\documentclass{llncs}
\usepackage{graphicx} 
\usepackage[hyphens]{url}
\usepackage[utf8]{inputenc}
\usepackage[english]{babel}
\usepackage{amsmath}
\usepackage{amsfonts}
\usepackage{amssymb}
\usepackage{booktabs}
\usepackage[amsmath,thref,thmmarks]{ntheorem}
\usepackage[algoruled]{algorithm2e}
\usepackage{algpseudocode}
\usepackage{tabularx}
\usepackage{tabu}
\usepackage[colorinlistoftodos,prependcaption,textsize=tiny]{todonotes}
\usepackage{xcolor-material}

\usepackage{listings}
\usepackage{listing-set}
\begin{document}

\mainmatter       % start of the contributions
%

% Oldies but Goldies
\title{Enriching Existing Test Collections with OXPath}

\author{
Philipp Schaer \and 
Mandy Neumann
}
\institute{
TH Köln (University of Applied Sciences), Cologne, Germany
\email{firstname.lastname@th-koeln.de}
}

\date{\today}

\maketitle

\begin{abstract}
% max 250 words
%1 motivation %2 problem statement %3 method %4 result/conclusion
Extending TREC-style test collections by incorporating external resources is a time consuming and challenging task. Making use of freely available web data requires technical skills to work with APIs or to create a web scraping program specifically tailored to the task at hand. We present a light-weight alternative that employs the web data extraction language OXPath to harvest data to be added to an existing test collection from web resources. We demonstrate this by creating an extended version of GIRT4 called GIRT4-XT with additional metadata fields harvested via OXPath from the social sciences portal Sowiport. This allows the re-use of this collection for other evaluation purposes like bibliometrics-enhanced retrieval.
The demonstrated method can be applied to a variety of similar scenarios and is not limited to extending existing collections but can also be used to create completely new ones with little effort.

\end{abstract}


\section{Introduction}
\label{sec:introduction}

Building TREC-style test collections for information retrieval evaluation is a costly activity. It involves at least three main tasks: (1) setting up an appropriate set of documents, (2) generating a list of topics (50 or even more, as suggested by Voorhees  \cite{Voorhees_2009}), and (3) obtaining relevance assessments (most of the time by employing domain experts or search specialists as assessors). All three tasks combined sum up and make the generation of new test collection or the redesign and extension of existing test collections a time consuming and challenging task. 
Generating a completely new test collections is the most complex scenario. Therefore we would like to focus on the enrichment of existing test collections, especially the set of documents. This would allow the reuse of documents, topics and relevance assessments while enabling the old test collection to be reused in other evaluation contexts like scholarly search or bibliometrics-enhanced retrieval \cite{Mayr_Scharnhorst_Larsen_Schaer_Mutschke_2014}.

Previous projects like EFIREval\footnote{\url{https://sites.google.com/site/ekanoulas/grants/EFIREval}} already focused on adapting test collections to new environments or incorporated richer information about the different retrieval scenarios and searchers’ activities but only few tried to augment the document collection, which is why we would like to focus on this desideratum. 
%We present an easy technical and methodological solution based on the web extraction language OXPath.

\noindent\textbf{Research question.} How can we enrich parts of existing test collections, like the document collection, by incorporating external resources like digital libraries or other freely available web data sets with as little effort as possible?

\noindent\textbf{Approach.} We propose a light-weight method for extending and augmenting the documents sets in test collections by incorporating the web extraction language OXPath. This language that derived from XPath is capable of extracting huge sets of information from large web corpora. It is used by scholarly literature portals like dblp to build up their data sets.

\noindent\textbf{Contributions.} We show the feasibility of our approach by extending the GIRT4 collection that was used in the Domain-Specific Track of CLEF with freely available data from the social sciences portal Sowiport\footnote{\url{http://sowiport.gesis.org}}. After harvesting the additional data we created an extended collection called GIRT4-XT that augments the original GIRT4 documents with additional attributes like ISSN codes. This way the rather old test collection that initially was used to do cross-lingual and domain-specific retrieval evaluations can be used for other evaluation purposes like bibliometrics-enhanced retrieval.

\section{Related Work}
\label{sec:relatedwork}

Re-using existing test collections for other purposes in general is not a new idea. Berendsen et al. \cite{Berendsen_Tsagkias_deRijke_Meij_2012} were using the previously mentioned GIRT collection to generate a so-called pseudo test collection that is automatically generated. The relatively spare data of the GIRT collection (content bearing metadata only being the title and a rather short abstract) comes with a rich set of annotations (see Table \ref{tab:girtmatrix}). These annotations were used to generate pseudo topics and relevance assessments. This pseudo test collection provided  training material for learning to rank methods. 

A similar approach was used by Roy, Ray, and Mitra \cite{Roy_Ray_Mitra_2016} who used the CiteSeerX collection to generate a test collection for citation recommendation services. They extracted the textual part of a citation context to form a query. The cited references were taken to be the relevant documents for that query. This way 2,826 queries were obtained but most queries (contexts) have only one relevant citation, making this test collection rather sparse.

Larsen and Lioma \cite{Larsen_Lioma_2016} described different strategies to generate a scholarly IDEAL test collection. While they came up with some new ideas and strategies of gathering and curating a document collection they rely on manually crafted topics and relevance assessments to complete the test collections. As they outline, the scholars that are the sources of topics and relevance assessments are notoriously busy, hard to engage and unlikely to be crowdsourced. They named INEX as a role model of community effort in collecting relevance judgments from its participants and encouraged to follow that road.
%There are more strategies to improve the topic generation and relevance assessment process. Kazai, Milic-Frayling, and Costella \cite{Kazai_Milic-Frayling_Costello_2009} proposed a method for the collective gathering of relevance assessments using a social game model to boost participants’ engagement. 

The reuse of document and test collections is common practice by adding new topics and relevance assessments or by transferring them to new application domains (e.g. from IR evaluation to recommender systems). Both approaches most often rely on manual work and judgments. Another approach for building up test collections was presented in the Social Book Search \cite{Koolen_Kazai_Preminger_Doucet_2013} track of CLEF. They built their task on top of the INEX Amazon/LibraryThing collection \cite{Beckers_Fuhr_Pharo_Nordlie_Fachry_2010} and enriched it with content from forum discussions on the LibraryThing website to extract topics and relevance assessments. This is a rather technical methodology to obtain this crucial part of a test collection which involved the generation of custom web crawlers for this single purpose.

%and previously INEX where crawled data from forum discussions were used to extract topics and relevance assessments.
%variante2
%This is a rather technical methodology to obtain this crucial part of a test collection where the Social Book Search team had to perform a targeted crawl on forum discussions which was not performed by the Social Book team itself but at University of Duisburg-Essen which involved the generation of custom web crawlers for this single purpose. 
%variante1
%The rather technical methodology to obtain this crucial part of a test collection was not performed by the Social Book team itself but at University of Duisburg-Essen which involved the generation of custom web crawlers for this single purpose. 
%\todo{reformulate}
%A rather technical methodology to obtain this crucial part of a test collection which was not performed by the Social Book team itself but at University of Duisburg-Essen which involved the generation of custom web crawlers for this single purpose. 

\section{Materials and Methods}
\label{sec:methods}
As suggested by some of the related work (e.g. Social Book Search), test collections can be created or enhanced with freely available web data. 
But web pages are meant to be displayed to a human user, as opposed to APIs that provide a means for software applications to gather the structured data that makes up the content of those web pages. Thus for compiling a corpus from web data, one would have to either have access to such an API, or work directly with the human-oriented HTML interface. The former would definitely require some programming/scripting skills, while the latter would either require extensive programming skills for scraping the web page content, or a lot of human effort to collect the desired information manually.
In the past, several attempts have been made to ease the process of acquiring web data for non-technical users, by providing web data extraction tools. 

OXPath is an open-source language focusing on deep web crawling that takes a declarative approach to the problem \cite{Furche_Gottlob_Grasso_Schallhart_Sellers_2013}. Based on the XML query language XPath, it enables the simulation of user interaction with a web page and the extraction of information in the course of these interactions. To achieve this, OXPath extends the capabilities of XPath with five new elements: (1) actions like clicking and form filling, (2) interactions with the visual appearance of a page, (3) means of identifying nodes by multiple relations, (4) extraction markers to yield hierarchical records of sought-after information, and (5) the Kleene star to enable navigation of paginated content. With these means, it is possible to craft an expression to harvest a lot of data with just a few lines of code. 

OXPath can be used e.g. for harvesting bibliographic metadata for digital libraries like dblp, as presented by Michels et al.  \cite{Michels_Fayzrakhmanov_Ley_Sallinger_Schenkel_2017}. In contrast to other tools made for extracting bulk data from web pages, OXPath proves to be particularly memory-efficient as shown by Furche et al. \cite{Furche_Gottlob_Grasso_Schallhart_Sellers_2013}.

Regarding document sets in test collections, OXPath can also be used to extract additional information from such digital libraries to extend the test collection with new attributes. Taking the social sciences portal Sowiport as an example, we created a light-weight OXPath wrapper that is able to harvest targeted information from a specific set of records and save the extracted data in a hierarchically structured form. Listing \ref{lst:sowiport} demonstrates a sample OXPath wrapper that is able to interact with the web page of Sowiport\footnote{Note that we replaced all German terms from the Sowiport portal with English equivalencies in this listing.}. It narrows down the list of presented items to those from a specific database (in this case the social science literature database SOLIS, that GIRT4 is based on) and navigates through the result list in a loop (lines 3--5). By clicking the title of each record element (line 7), the element's detail view is opened where additional data can be found. For example, in lines 8--10 the editor field is located in the page and each listed editor extracted separately. In a similar vein, the acquisition id (``Acquis. id'') is extracted from a different location on the same page (lines 11--13). The extracted data is hierarchical in nature and can be serialized e.g. in XML or CSV format for further processing.

%The extracted data was then used to enhance the GIRT4 collection with additional attributes as described in the following section.

%
\begin{lstlisting}[%
float=t,
linewidth=\columnwidth,
language=oxpath,
caption={Sample OXPath wrapper for harvesting the SOLIS database of Sowiport and extracting title, language and published attributes for each result.},
label=lst:sowiport,
]
doc('http://sowiport.gesis.org/')
  /descendant::field()[2]/{click /}
    //div[@id='facets']//a[contains(.,'SOLIS')]/{click /}
    //*[@id='limit']/option[contains(., '100')]/{click /}
    /(//*[contains(@title, 'next')]/{click /})*
     //*[contains(@class,'record')]:<record>
      [.//a[@class~='title'][./b:<title=normalize-space(.)>]/{click /}
        /. [? .//*[@id='detailed_view_metadata']//table//td
         [ ./preceding-sibling::td[contains(.,'Editor:')]]//a
           :<editor=normalize-space(.)>]
        [? .//*[@class='recordsubcontent']//table//td
         [ ./preceding-sibling::th[contains(.,'Database:')]]
           :<id=substring-after(normalize-space(.), 'Acquis. id: ')>]
      ]
\end{lstlisting}

\section{Results}
\label{sec:results}
%Mandy
By harvesting additional data from the SOLIS database in Sowiport using a relatively simple declarative expression, we were able to extend the original GIRT4 data with additional information, such as ISSN/ISBN codes or editor, publisher and location information (see Table \ref{tab:girtmatrix} for an overview). The items from the GIRT collection were matched with the harvested data via their id which was both present in the harvested SOLIS data (acquisition id) and the GIRT4 data set (DOCID without the GIRT prefix). 

Of a total of 151.319 documents in GIRT4 we extended 135.214 documents with data from SOLIS/Sowiport. Note that only the documents on social science literature were extended while the social science project descriptions also included in GIRT were ignored. The new test collection is called GIRT4-XT and includes a total of six new metadata fields that were not included in the original data set (editor, ISSN, ISBN, location, publisher, and page numbers). Some of the SOLIS records include links to full texts but as most of them are behind publisher pay walls we were not able to extract them.

\begin{table}[t]
\caption{
    Overview on the included fields of the original GIRT4 corpus, the available SOLIS data from the Sowiport portal and the combined GIRT4-XT corpus. Three different states are marked in the table: \textendash{}
    = field data not available ; $\circ$ = available in unstructured form; $\bullet$ = available in structured form.
}
\begin{tabularx}{1\textwidth}{l@{\quad} X@{\quad} X@{\quad} X@{\quad} X@{\quad} X@{\quad} X@{\quad} X@{\quad} X@{\quad} X@{\quad} X@{\quad} X@{\quad} X@{\quad} X@{\quad} X@{\quad} X@{\quad} X@{\quad} X@{\quad} X}
\toprule 
Corpus & \rotatebox{90}{id} & \rotatebox{90}{author} & \rotatebox{90}{editor} & \rotatebox{90}{title} & \rotatebox{90}{source} & \rotatebox{90}{issn} & \rotatebox{90}{isbn} & \rotatebox{90}{pubyear} & \rotatebox{90}{keywords} & \rotatebox{90}{class.} & \rotatebox{90}{abstract} & \rotatebox{90}{full text} & \rotatebox{90}{method} & \rotatebox{90}{location} & \rotatebox{90}{publisher} & \rotatebox{90}{pages} & \rotatebox{90}{language} & \rotatebox{90}{country} \\
\midrule
GIRT4 & $\bullet$ & $\bullet$ & \textendash{} & $\circ$ & $\circ$ & \textendash{} & \textendash{} & $\bullet$ & $\bullet$ & $\bullet$ & $\bullet$ & \textendash{} & $\bullet$ & \textendash{} & \textendash{} & \textendash{} & $\bullet$ & $\bullet$\tabularnewline
\addlinespace
SOLIS & $\bullet$ & $\bullet$ & $\bullet$ & $\circ$ & $\bullet$ & $\bullet$ & $\bullet$ & $\bullet$ & $\bullet$ & $\bullet$ & $\bullet$ & $\circ$ & $\bullet$ & $\circ$ & $\circ$ & $\bullet$ & $\bullet$ & $\bullet$\tabularnewline
\addlinespace
GIRT4-XT & $\bullet$ & $\bullet$ & $\bullet$ & $\circ$ & $\circ$ & $\bullet$ & $\bullet$ & $\bullet$ & $\bullet$ & $\bullet$ & $\bullet$ & \textendash{} & $\bullet$ & $\circ$ & $\circ$ & $\bullet$ & $\bullet$ & $\bullet$\tabularnewline
\bottomrule
\addlinespace
\end{tabularx}
\label{tab:girtmatrix}
\end{table}

\section{Discussion and Conclusion}
\label{sec:discussion}
We showed how to extend and enrich existing information retrieval test collections by harvesting freely available metadata from digital library systems by employing the web extraction language OXPath. This method allows us to reuse existing test collections (especially their topics and relevance assessments) in different domains by adding new metadata to the existing documents in the collection. 

We demonstrated the feasibility of the process by extending GIRT4 with additional document annotations like editor names, ISSN codes of the related journal or page numbers. This way new kinds of experiments are possible like those discussed in the bibliometrics-enhanced IR community, but the proposed methods and techniques are not limited to this domain. Another use case for our test collection enrichment strategy might be the TREC Genomics Track test collections\footnote{\url{http://skynet.ohsu.edu/trec-gen/}}. As suggested by Larsen and Lioma \cite{Larsen_Lioma_2016} these collections can be augmented by references extracted from PubMed, a scenario more than suitable for OXPath.

The proposed approach heavily relies on the usage of OXPath as it is an easy-to-learn, light-weight, and all-in-one rapid development technology to gather the additional (meta-)data from web resources like digital libraries. Although the advantages outweigh the disadvantages we would like to point out some shortcomings of OXPath that have to be considered. First of all OXPath is not tuned for speed which results in rather moderate processing times. Internally the whole web page has to be rendered and processed to allow a human-comparable extraction mechanism. When processing many hundred thousand web pages the harvesting process can take many days. There are ways to distribute the whole process on parallel threads but this is not a built-in feature. Another point is that there are relatively few tools to support the development process\footnote{We developed an extension for the text editor Atom ourselves, see \url{https://atom.io/packages/language-oxpath}}. In spite of these limitations, OXPath is still a powerful and useful tool for harvesting semi-structured data from web resources. %In particular, compared with other approaches, ours is neither dependent on the existence of an API, nor does it require high-level programming skills.

In the future, we want to employ OXPath not only for the enhancement of existing test collections, but also for the creation of completely new ones, were all the data necessary should be extracted from web resources. One of our role models for this is the Social Book Search collection.

\subsubsection{Acknowledgements.}
This work was supported by Deutsche Forschungsgemeinschaft (DFG), grant no. SCHA 1961/1-2.

\bibliographystyle{splncs03}
\bibliography{oxpathtestcol}

\end{document}